\documentclass[a4paper,11pt]{article}
\pdfoutput=1 
\usepackage{aas_macros,braket}
\usepackage{jcappub,natbib} 
\usepackage[T1]{fontenc} 

\title{General modal estimation for cross-bispectra}

\author[a]{Maresuke Shiraishi,}
\author[b,c,d]{Michele Liguori,}
\author[e]{James R. Fergusson,}
\author[e]{and E.P.S. Shellard}

\affiliation[a]{Department of General Education, National Institute of Technology, Kagawa College, 355 Chokushi-cho, Takamatsu, Kagawa 761-8058, Japan}
\affiliation[b]{Dipartimento di Fisica e Astronomia ``G. Galilei'', Universit\`a degli Studi di Padova, via Marzolo 8, I-35131, Padova, Italy}
\affiliation[c]{INFN, Sezione di Padova, via Marzolo 8, I-35131, Padova, Italy}
\affiliation[d]{INAF-Osservatorio Astronomico di Padova, Vicolo dell'OSservatorio 5, I-35122 Padova, Italy}
\affiliation[e]{Centre for Theoretical Cosmology, DAMTP, University of Cambridge, Wilberforce Road, Cambridge, CB3 0WA United Kingdom}

\emailAdd{shiraishi-m@t.kagawa-nct.ac.jp}
\emailAdd{michele.liguori@pd.infn.it}
\emailAdd{J.Fergusson@damtp.cam.ac.uk}
\emailAdd{epss@damtp.cam.ac.uk}

\abstract{
  We describe a fast, optimal estimator for measuring angular bispectra between two correlated weakly non-Gaussian fields ($Y$ and $Z$) from observational datasets, based on a separable modal bispectrum expansion. Our methodology is applicable to (1) any shape of the input theoretical bispectrum templates (factorizable or not), (2) both even and odd $\ell_1 + \ell_2 + \ell_3$ multipole domains and (3) both amplitude ($f_{\rm NL})$ bispectrum estimation and full bispectrum reconstruction, considering either joint estimation of ($YYY$, $ZZZ$, $YYZ$ and $ZZY$) shapes, or independent estimation of auto-bispectra ($YYY$ or $ZZZ$) and cross-bispectra ($YYZ$ or $ZZY$); hence, it has quite high versatility. The methodology described here was implemented and used for the official analysis of temperature and polarization cosmic microwave background maps from the {\it Planck} satellite.}

\begin{document}



\maketitle
\flushbottom

\section{Introduction}

Non-Gaussianity (NG) in cosmic observables is an important indicator of the statistical and physical properties of the cosmological fluctuation field, carrying information both about primordial perturbations and later processes, such as reionization and structure formation. 

A typical and useful approach to test cosmological NG is the study of beyond two-point correlation functions. Among them, the bispectrum, i.e., the Fourier transform of the three-point correlation, often plays a special role. For example, if we focus on the primordial fluctuation field generated by inflation, the bispectrum of Cosmic Microwave Background (CMB) anisotropies is generally the most powerful test of models beyond standard single-field slow roll. 
Stringent constraints on primordial NG have been initially obtained using the auto-bispectrum of the temperature field ($T$) in WMAP and {\it Planck} data \cite{Bennett:2012zja,Ade:2013ydc}. These were further refined by including the CMB polarization field in the latest {\it Planck} data analysis \cite{Ade:2015ava}. 
 Optimal estimation of primordial NG from CMB temperature and polarization data requires constraining jointly two auto-bispectra ($TTT$ and $EEE$) and two mixed ones ($TTE$ and $EET$). There are many other interesting examples, in the literature, of cross-bispectra containing relevant cosmological information, see e.g. refs.~\cite{Kamionkowski:2010rb,Shiraishi:2011st,Shiraishi:2013vha,Shiraishi:2013kxa,Meerburg:2016ecv,Domenech:2017kno,Bartolo:2018elp} for the correlators containing CMB B-mode polarization and refs.~\cite{Bartolo:2015fqz,Shiraishi:2016hjd} for those with CMB spectral distortion anisotropies. 

A main purpose of this paper is therefore to establish a general methodology for optimal estimation of angular bispectra composed of two correlated weakly NG fields. In order for it to be as general as possible, as well as to encompass all the examples mentioned in the list above, we want our approach to be able to deal with both parity-odd and parity-even bispectra, eventually written in non-separable form (i.e. angular bispectra that cannot be explicitly written as factorized products of the three multipole moments $\ell_1$, $\ell_2$ and $\ell_3$; these models pose significant computational challenges, as explained in detail in the main text below). We tackle these problems by specifically resorting to a modal estimation approach, in which a general bispectrum is written as a linear combination 
of factorizable basis templates, and the latter are fit to the data. The work developed here will therefore provide a generalization of the modal bispectrum estimation pipeline developed for auto-bispectra in refs.~\cite{Fergusson:2006pr,Fergusson:2009nv,Fergusson:2010dm,Shiraishi:2014roa,Shiraishi:2014ila}.  
Another extension of the modal estimation approach, aimed at joint estimation of temperature and polarization CMB bispectra, is developed and discussed in 
ref.~\cite{Fergusson:2014gea}. There, the author orthogonalizes the input bispectrum vector $(B_{\ell_1 \ell_2 \ell_3}^{YYY}, B_{\ell_1 \ell_2 \ell_3}^{ZZZ}, B_{\ell_1 \ell_2 \ell_3}^{YYZ}, B_{\ell_1 \ell_2 \ell_3}^{ZZY})$ (where $Y$ and $Z$ are two general weakly NG random fields; in the case of CMB {\it Planck} analysis, $Y$ and $Z$ represent the temperature and E-mode polarization fields). After rotation, the problem is reduced to that of estimating four independent auto-bispectra in the new variables. 

In this paper we take a different approach, namely we directly decompose the original bispectra, without rotation. As we show in the main text, this approach can also yield a factorized form of the estimator \eqref{eq:estimator_2D}, allowing for fast bispectrum estimation, although with a slightly worse performance than the orthogonalized method. On the other hand, the approach discussed here allows us to very easily obtain a fast, optimal modal estimator for the (physical, un-rotated) mixed bispectra ($YYZ$ or $ZZY$) {\em alone}. This can actually be applied to validation and consistency tests in a joint bispectrum analysis, as well as to the estimation of specific cross-bispectra, such as, for example, CMB $TTB$ bispectra mentioned above, which have recently received significant attention in the literature. The methodology we develop in this paper is therefore different and complementary to the orthogonalized modal estimator of ref.~\cite{Fergusson:2014gea}. We note that both methods have already been implemented and used for a large range of applications in the {\it Planck} data analysis. They are dubbed, respectively, Modal1 (estimator discussed in this paper) and Modal2 (orthogonalized estimator) methods in the {\it Planck} 2015 analysis of primordial NG \cite{Ade:2015ava}.

This paper is organized as follows. In 
section~\ref{sec:def}, we briefly recall some general results in the theory of bispectrum estimation, and report the factorized forms of the optimal angular bispectrum estimator \eqref{eq:estimator_2D}. In section~\ref{sec:even} we describe our modal decomposition methodology for even-parity mixed bispectra  \cite{Fergusson:2006pr,Fergusson:2009nv,Fergusson:2010dm}. In section~\ref{sec:odd}, we generalize the method to odd-parity mixed bispectra. In section~\ref{sec:mixed}, we discuss how our modal approach can be directly used not only to build fast estimators of bispectrum amplitudes ($f_{\rm NL}$) from a joint analysis of the four auto- and cross-bispectra, but also to independently reconstruct their shapes. In section \ref{sec:tests} we discuss the implementation and validation of our method. Our conclusions are then summarized in the final section.

\section{Optimal $f_{\rm NL}$ estimation and non-Gaussian map generation}\label{sec:def}

We consider a pair of weakly NG fields, $Y$ and $Z$, and define their angular auto- and cross-bispectra as follows:
\begin{eqnarray}
  \begin{split}
  \Braket{a_{\ell_1 m_1}^Y a_{\ell_2 m_2}^Y a_{\ell_3 m_3}^Y}
  &\equiv f_{\rm NL} \, B_{\ell_1 \ell_2 \ell_3}^{YYY}
  \left(
  \begin{array}{ccc}
  \ell_1 & \ell_2 & \ell_3 \\
  m_1 & m_2 & m_3 
  \end{array}
  \right) , \\
    \Braket{a_{\ell_1 m_1}^{Z} a_{\ell_2 m_2}^{Z} a_{\ell_3 m_3}^{Z}}
  &\equiv f_{\rm NL} \, B_{\ell_1 \ell_2 \ell_3}^{ZZZ}
  \left(
  \begin{array}{ccc}
  \ell_1 & \ell_2 & \ell_3 \\
  m_1 & m_2 & m_3 
  \end{array}
  \right) , \\
  \Braket{a_{\ell_1 m_1}^{Y} a_{\ell_2 m_2}^{Y} a_{\ell_3 m_3}^{Z}}
  &\equiv f_{\rm NL} \, B_{\ell_1 \ell_2 \ell_3}^{YYZ}
  \left(
  \begin{array}{ccc}
  \ell_1 & \ell_2 & \ell_3 \\
  m_1 & m_2 & m_3 
  \end{array}
  \right) , \\
    \Braket{a_{\ell_1 m_1}^{Z} a_{\ell_2 m_2}^{Z} a_{\ell_3 m_3}^{Y}}
  &\equiv f_{\rm NL} \, B_{\ell_1 \ell_2 \ell_3}^{ZZY}
  \left(
  \begin{array}{ccc}
  \ell_1 & \ell_2 & \ell_3 \\
  m_1 & m_2 & m_3 
  \end{array}
  \right) ,
  \end{split} \label{eq:fNL_def}
\end{eqnarray}
where $f_{\rm NL}$ is the usual NG-amplitude parameter, defining the strength of the NG signal. For example, in CMB studies of primordial NG, $Y$ and $Z$ can be any two between the temperature, E-mode polarization and B-mode polarization fields. Here we note that while in this paper we demonstrate the method for fields on a 2 sphere, with the specific example of the CMB, the method can be trivially extended to any general set of $M$ correlated fields in $N$-dimensions.%
\footnote{For example, a complete CMB bispectrum estimation requires the method for $M = 3$ (in order to deal with $TEB$ correlator). It can also be achieved by a simple extension of our $M=2$ method.}

We start here by focusing on the problem of optimal $f_{\rm NL}$ estimation, using all the available auto- and cross-bispectra between $Y$ and $Z$. In other words, we consider the problem of estimating the bispectrum amplitude by selecting a specific NG model and fitting the theoretically predicted shape to the data. The optimal joint-estimator for $f_{\rm NL}$ was originally described in refs.~\cite{Babich:2004yc,Yadav:2007rk} and it can be written as:%
\footnote{
This formula is perturbatively derived assuming the weakness of NG and hence is not suitable for analyzing strong NG templates such as some secondary bispectra and bispectra estimated using fields in the nonlinear regime. Moreover, this is justified only when the covariance matrix is approximately diagonalized. Such a situation has actually been realized in the {\it Planck} 2015 data analysis owing to a recursive inpainting process \cite{Ade:2015ava}. We explore this case in the following and derive the fast modal estimator~\eqref{eq:estimator_2D_mod}. If discarding the diagonal covariance matrix approximation, our derivation will be slightly complicated by replacing $\sum_{X'} (C_\ell^{-1})^{X X'} a_{\ell m}^{X'}$ with $\sum_{X'} \sum_{\ell' m'} ({\cal C}^{-1})_{\ell m \ell' m'}^{X X'} a_{\ell' m'}^{X'}$, where ${\cal C}_{\ell m \ell' m'}^{X X'} \equiv \Braket{a_{\ell m}^{X *} a_{\ell' m'}^{X'}}$. It is then expected that the inverse of the covariance matrix of the $\beta$ coefficients are inserted into the estimator~\eqref{eq:estimator_2D_mod} as Ref.~\cite{Fergusson:2014gea} reported.}
\begin{equation}
  {\cal E}
= \frac{1}{6F}
\sum_{X_i X_i'} \sum_{\ell_i}
(-1)^{\ell_1 + \ell_2 + \ell_3} 
B_{\ell_1 \ell_2 \ell_3}^{X_1 X_2 X_3} 
(C_{\ell_1}^{-1})^{X_1 X_1'} 
(C_{\ell_2}^{-1})^{X_2 X_2'} (C_{\ell_3}^{-1})^{X_3 X_3'} 
{\cal B}_{\ell_1 \ell_2 \ell_3}^{X_1' X_2' X_3'} , \label{eq:estimator_2D}
\end{equation}
where $X_i$ and $X_i'$ span over $Y$ and $Z$, and $(C_\ell^{-1})^{X X'}$ is the $(X,X')$ element of the inverse 2D angular power spectrum matrix $\left(
\begin{smallmatrix}
  C_\ell^{YY} & C_\ell^{YZ}  \\
  C_\ell^{YZ} & C_\ell^{ZZ}
  \end{smallmatrix}
  \right)^{-1}$, ${\cal B}_{\ell_1 \ell_2 \ell_3}^{X_1' X_2' X_3'}$ are the observed angle-averaged bispectra given by
\begin{equation}
  {\cal B}_{\ell_1 \ell_2 \ell_3}^{X_1' X_2' X_3'}
  \equiv
  \sum_{m_i} \left(
  \begin{array}{ccc}
  \ell_1 & \ell_2 & \ell_3 \\
  m_1 & m_2 & m_3 
  \end{array}
  \right)
  \left[ 
    a_{\ell_1 m_1}^{X_1'} a_{\ell_2 m_2}^{X_2'} a_{\ell_3 m_3}^{X_3'}
    - \left(a_{\ell_1 m_1}^{X_1'}
    \Braket{a_{\ell_2 m_2}^{X_2'} a_{\ell_3 m_3}^{X_3'}}
+ 2 \ {\rm perms} \right)
\right] ,
\end{equation}
and $F$ is the Fisher matrix:
\begin{equation}
  F = \frac{1}{6} \sum_{X_i X_i'} \sum_{\ell_i}
  (-1)^{\ell_1 + \ell_2 + \ell_3} 
  B_{\ell_1 \ell_2 \ell_3}^{X_1 X_2 X_3}
(C_{\ell_1}^{-1})^{X_1 X_1'} (C_{\ell_2}^{-1})^{X_2 X_2'} (C_{\ell_3}^{-1})^{X_3 X_3'} 
B_{\ell_1 \ell_2 \ell_3}^{X_1' X_2' X_3'} . \label{eq:F_2D}
\end{equation}
Note that we reported here explicitly the $(-1)^{\ell_1 + \ell_2 + \ell_3}$ term, which is usually dropped in the literature, since we are interesting here in a general estimator dealing with both even and odd $\ell_1 + \ell_2 + \ell_3$ configurations.

Our goal in the next few sections will be that of defining a general modal approach to implement this optimal joint-estimator of $f_{\rm NL}$. We will however point out 
in section \ref{sec:mixed} that, with some simple, direct rearrangement of the output and no further computational cost, the same modal estimation pipeline also produces {\em model-independent} 
reconstruction of the full auto- and cross-bispectra from the data, as well as fast and optimal $f_{\rm NL}$ measurements, obtained by fitting $YYZ$, $ZZY$, $YYY$ and $ZZZ$ independently, and not jointly with all other combinations.

For estimator testing purposes, it is useful to have a procedure to generate NG maps with given power spectrum and bispectrum. A fast algorithm achieving this goal was originally proposed in ref.~\cite{Smith:2006ud} for auto-spectra, and generalized to multiple fields with non-vanishing cross-spectra in ref.~\cite{Fergusson:2014gea}. 

A set of multipoles $(a_{\ell m}^{Y}, a_{\ell m}^{Z})$, with specified two-point correlation $\braket{a_{\ell_1 m_1}^{X_1} a_{\ell_2 m_2}^{X_2}} = C_{\ell_1}^{X_1 X_2} (-1)^{m_1} \delta_{\ell_1, \ell_2} \delta_{m_1, -m_2} $ and the 3-point one~\eqref{eq:fNL_def} can be generated according to the following prescription: 
\begin{equation}
  \left(
\begin{array}{c}
  a_{\ell m}^Y \\
  a_{\ell m}^Z
  \end{array}
  \right)
  = L_\ell^{-1}
  \left(
\begin{array}{c}
  g_{\ell m}^Y + A_{\ell m}^{YYY} \\
  g_{\ell m}^Z
+ 3 A_{\ell m}^{ZYY} + 3 A_{\ell m}^{ZZY} + A_{\ell m}^{ZZZ}
  \end{array}
  \right), 
\end{equation}
where $g_{\ell m}^X$ are orthonormal Gaussian realizations obeying
$\braket{g_{\ell_1 m_1}^{X_1} g_{\ell_2 m_2}^{X_2}} = (-1)^{m_1} \delta_{\ell_1, \ell_2} \delta_{m_1, -m_2} \delta_{X_1, X_2}$
and
\begin{eqnarray}
  L_\ell
  &\equiv& 
\left(
  \begin{array}{ccc}
    L_\ell^{YY} & L_\ell^{YZ} \\ 
    L_\ell^{ZY} & L_\ell^{ZZ} 
  \end{array}
  \right)
= 
\left(
  \begin{array}{ccc}
  \frac{1}{\sqrt{C_\ell^{YY}}} & 0 \\
  \frac{- C_\ell^{YZ}}{\sqrt{C_\ell^{YY}}\sqrt{C_\ell^{YY} C_\ell^{ZZ} - (C_\ell^{YZ})^2} } & 
 \frac{C_\ell^{YY}}{\sqrt{C_\ell^{YY}}\sqrt{C_\ell^{YY} C_\ell^{ZZ} - (C_\ell^{YZ})^2} }
  \end{array}
  \right), \label{eq:L_def} \\
  A_{\ell_1 m_1}^{X_1 X_2 X_3} 
&=& 
\frac{1}{6}
 \sum_{\ell_2 m_2 \ell_3 m_3} 
g_{\ell_2 m_2}^{X_2 *} g_{\ell_3 m_3}^{X_3 *}
\sum_{X_i'}
L_{\ell_1}^{X_1 X_1'} L_{\ell_2}^{X_2 X_2'} L_{\ell_3}^{X_3 X_3'}
 B_{\ell_1 \ell_2 \ell_3}^{X_1' X_2' X_3'}
\left(
  \begin{array}{ccc}
  \ell_1 & \ell_2 & \ell_3 \\
  m_1 & m_2 & m_3
  \end{array}
 \right). \label{eq:Alm_sim}
\end{eqnarray}
We will show later how the generation of NG maps can be made computationally efficient using the modal approach.

\section{Modal estimator for parity-even bispectra}\label{sec:even}

The well-known, general problem with formula~\eqref{eq:estimator_2D} is that the brute-force computation of this estimator demands ${\cal O}(\ell_{\rm max}^5)$ operations, where $\ell_{\rm max}$ is the maximum number of available multipoles. When $\ell_{\rm max} \gtrsim 10^3$ (comparable to {\it Planck} resolution, in the CMB case \cite{Ade:2013ydc,Ade:2015ava}), this becomes a completely unfeasible task. In order to reduce the computational cost in bispectrum estimation, a variety of computational methodologies have been developed. We here work with one of them, the so-called Modal methodology, originally developed in refs.~\cite{Fergusson:2009nv,Fergusson:2010dm}. For other approaches, the reader can refer to, e.g., ref.~\cite{Komatsu:2003iq} (KSW estimator), ref.~\cite{Bucher:2015ura} (Binned estimator), and ref.~\cite{MartinezGonzalez:2001gu} (Wavelet method).%
\footnote{
In fact both the binned and wavelet approaches, and indeed all optimal methods for bispectrum estimation, can be encompassed by the modal approach simply by changing the set of basis functions to the appropriate form, e.g. top hat functions in $\ell$-space for bins or the harmonic transform of the wavelet for wavelet approaches.  This is a reflection of the fact that the only solution to evaluating the optimal bispectrum estimator is to represent the bispectrum in a separable form and the modal method is simply a generic procedure for doing this.
}

It was first realized by the authors of ref.~\cite{Komatsu:2003iq} that the number of operations required for NG estimation can be reduced to ${\cal O}(\ell_{\rm max}^3)$, with a massive ${\cal O}(\ell_{\rm max}^2)$ gain, if the theoretical bispectrum template in input is written in separable form. This is however often not the case. The point of the modal methodology is that of tightly approximating the input bispectrum, by expanding it into finite sets of separable eigenfunctions in bispectrum space. This approach was initially developed for parity-even, temperature-only, CMB bispectrum measurements \cite{Fergusson:2009nv,Fergusson:2010dm} and was later extended to the parity-odd temperature-only case \cite{Shiraishi:2014roa,Shiraishi:2014ila}. In the following, we consider the direct extension of the modal methodology for auto-bispectra, with the goal to perform joint estimation of CMB temperature and polarization bispectra, or any other general combination of auto- and cross-bispectra.

In this section, we consider parity-even angle-averaged bispectra, the nonvanishing signal of which is confined to
\begin{equation}
  \ell_1 + \ell_2 + \ell_3 = {\rm even}, \ \ \
  |\ell_1 - \ell_2| \leq \ell_3 \leq |\ell_1 + \ell_2| . \label{eq:domain_even}
\end{equation}
Such bispectra can be expressed as
\begin{equation}
  B_{(e) \ell_1 \ell_2 \ell_3}^{X_1 X_2 X_3} 
\equiv h_{\ell_1 \ell_2 \ell_3} b_{(e) \ell_1 \ell_2 \ell_3}^{X_1 X_2 X_3} ,  \label{eq:Blll_even}
\end{equation}
where a geometrical factor:
\begin{equation}
h_{\ell_1 \ell_2 \ell_3} 
\equiv \sqrt{\frac{(2\ell_1 + 1)(2\ell_2 + 1)(2\ell_3 + 1)}{4\pi}}
\left(
  \begin{array}{ccc}
  \ell_1 & \ell_2 & \ell_3 \\
  0 & 0 & 0 
  \end{array}
 \right),
\end{equation}
filters the multipoles satisfying eq.~\eqref{eq:domain_even}. The reduced bispectra $b_{(e) \ell_1 \ell_2 \ell_3}^{X_1 X_2 X_3}$ contain all physical information.

\subsection{Modal decomposition}

Our initial goal is that of finding a factorizable representation for a general mixed reduced bispectrum, $b_{(e) \ell_1 \ell_2 \ell_3}^{X_1 X_2 X_3}$. 
This is done here via a straightforward generalization of the modal approach for auto-bispectra, described in ref.~\cite{Fergusson:2009nv}.
To start with, we consider a set of factorizable basis templates, $Q_{ijk}^{X_1 X_2 X_3}$, defined as:
\begin{equation}
Q_{ijk}^{X_1 X_2 X_3} (\ell_1, \ell_2, \ell_3)  \equiv q_{[i}^{X_1}(\ell_1) q_j^{X_2}(\ell_2) q_{k]}^{X_3}(\ell_3) ,
\end{equation}
where $q_i^X(\ell)$ is a general set of eigenfunctions, which can be in principle arbitrarily chosen. Different choices can however have a strong impact on the rate of convergence for specific input theoretical shapes. Common choices are a polynomial set or a set of oscillatory trigonometric functions; see refs.~\cite{Fergusson:2009nv,Fergusson:2014gea} for more details on this issue. The notation $[i, j, k]$ represents permutations over the indices $i$, $j$ and $k$, in order to obtain suitably symmetrized representations, for either auto- or cross-bispectra, i.e.
\begin{eqnarray}
  \begin{split}
    q_{[i}^{Y}(\ell_1) q_j^{Y}(\ell_2) q_{k]}^{Y}(\ell_3)
    &=
    \frac{1}{6} \left[ 
      q_i^{Y}(\ell_1) q_j^{Y}(\ell_2) q_k^{Y}(\ell_3) + 5~{\rm perms~in~}i,j,k \right] , \\
    q_{[i}^{Y}(\ell_1) q_j^{Y}(\ell_2) q_{k]}^{Z}(\ell_3)
    &= 
    \frac{1}{2}\left[   q_i^{Y}(\ell_1) q_j^{Y}(\ell_2) + q_j^Y(\ell_1) q_i^Y(\ell_2) \right] 
    q_k^{Z}(\ell_3) . 
  \end{split}
  \label{eq:expansion}
\end{eqnarray}
Now we want to approximate the reduced bispectrum in a way that ensures the minimum error is introduced to the estimator~\eqref{eq:estimator_2D}.  To do this we must decompose the signal-to-noise weighted reduced bispectrum which we can easily determine from the Fisher matrix \eqref{eq:F_2D}, to be:
\begin{equation}
\frac{h_{\ell_1 \ell_2 \ell_3}  b_{(e) \ell_1 \ell_2 \ell_3}^{X_1 X_2 X_3}  }{\sqrt{C_{\ell_1}^{X_1 X_1} C_{\ell_2}^{X_2 X_2} C_{\ell_3}^{X_3 X_3}} } \, . 
\end{equation}
Unfortunately $h_{\ell_1 \ell_2 \ell_3}$ is not easily separable and so we instead mimic it with the good separable approximation $v_{\ell_1} v_{\ell_2} v_{\ell_3}$ where $v_{\ell} = (2 \ell + 1)^{1/6}$; see again \cite{Fergusson:2009nv} for a more detailed discussion on this specific point. We can now decompose the approximately signal-to-noise weighted bispectrum into our separable modal basis $Q_{ijk}^{X_1 X_2 X_3}(\ell_1, \ell_2, \ell_3)$, as
\begin{equation}
\frac{v_{\ell_1} v_{\ell_2} v_{\ell_3} b_{(e) \ell_1 \ell_2 \ell_3}^{X_1 X_2 X_3}  }{\sqrt{C_{\ell_1}^{X_1 X_1} C_{\ell_2}^{X_2 X_2} C_{\ell_3}^{X_3 X_3}} }  
= \sum_{ijk} \alpha_{ijk}^{X_1 X_2 X_3} Q_{ijk}^{X_1 X_2 X_3}(\ell_1, \ell_2, \ell_3) . \label{eq:decompQ_even} 
\end{equation}

The level of convergence of the expansion to the original shape is measured in terms of the following shape correlator, which naturally defines a (parity-even) inner product in bispectrum space:
\begin{equation} 
\gamma_{np}^{X_1 X_2 X_3} \equiv \Braket{Q_{n}^{X_1 X_2 X_3}(\ell_1, \ell_2, \ell_3)  \, Q_{p}^{X_1 X_2 X_3}(\ell_1, \ell_2, \ell_3) }_{e} , \label{eq:gamma_even} 
\end{equation}
where
\begin{equation}
  \Braket{ f_{\ell_1 \ell_2 \ell_3}^{X_1 X_2 X_3 X_1' X_2' X_3'} }_e \equiv \sum_{\ell_i} \left( \frac{h_{\ell_1 \ell_2 \ell_3}}{v_{\ell_1} v_{\ell_2} v_{\ell_3}} \right)^2
f_{\ell_1 \ell_2 \ell_3}^{X_1 X_2 X_3 X_1' X_2' X_3'} , \label{eq:prod_even_def} 
\end{equation}
and the squared term in brackets corrects for using our approximation to $h_{\ell_1 \ell_2 \ell_3}$ in the decomposition. Note also that in this last expression we have labeled the triples $ijk$ in $Q_{ijk}^{X_1 X_2 X_3}$ by means of a single index $n$. Using this, the modal coefficients are expressed as
\begin{equation}
\alpha_n^{X_1 X_2 X_3} = \sum_p \left(\gamma^{X_1 X_2 X_2}\right)_{np}^{-1} 
\Braket{ \frac{v_{\ell_1} v_{\ell_2} v_{\ell_3} b_{(e) \ell_1 \ell_2 \ell_3}^{X_1 X_2 X_3} }{ \sqrt{C_{\ell_1}^{X_1 X_1} C_{\ell_2}^{X_2 X_2} C_{\ell_3}^{X_3 X_3}}} \, Q_p^{X_1 X_2 X_3} (\ell_1,\ell_2,\ell_3) }_{e} . 
\end{equation} 

Note that, in general, $\gamma_{np}^{X_1 X_2 X_3} \neq \delta_{np}$, and the $Q_n^{X_1 X_2 X_3}$ templates then form a non-orthonormal basis. 
We can however always produce an orthonormal basis $R_n$, starting from $Q_n$, via a suitable rotation:
\begin{equation}
R_n^{X_1 X_2 X_3}(\ell_1, \ell_2, \ell_3)  = \sum_p \lambda_{np}^{X_1 X_2 X_3}  Q_p^{X_1 X_2 X_3}(\ell_1, \ell_2, \ell_3) . \label{eq:Q2R}
\end{equation}
In the formula above, the lower triangular, rotation matrix $\lambda_{np}^{X_1 X_2 X_3}$ is related to $\gamma_{np}^{X_1 X_2 X_3}$ via a Cholesky decomposition:
\begin{equation}
\left( \gamma^{X_1 X_2 X_3} \right)^{-1} = \left( \lambda^{X_1 X_2 X_3} \right)^\top \, \lambda^{X_1 X_2 X_3} . 
\end{equation}
In total analogy with what done above, one can expand the input bispectrum replacing the $Q_{n}^{X_1 X_2 X_3}$ basis templates with their orthonormal counterparts $R_n^{X_1 X_2 X_3}$, extracting the new expansion coefficients $\alpha_n^{R X_1 X_2 X_3}$ via the inner product $\braket{ b, R }_e$. The relation between $\alpha_n^{X_1 X_2 X_3}$ and $\alpha_n^{R X_1 X_2 X_3}$ is then obtained as
\begin{equation}
  \alpha_n^{X_1 X_2 X_3} = \sum_p \lambda^{X_1 X_2 X_3}_{pn} \alpha_p^{R X_1 X_2 X_3} . \label{eq:aR2aQ} 
\end{equation}
It must however be noted that the orthonormal $R$ templates are generally not separable. Since separability is crucial for fast estimation, the actual estimator employs the non-orthonormal $Q$-basis, and the rotation is then often applied to the measured template amplitudes, in order to simplify and make the interpretation of the results more transparent.

In the modal decomposition process, supposing that the input bispectrum, $b_{(e) \ell_1 \ell_2 \ell_3}^{X_1 X_2 X_3}$, is precomputed, the most time-consuming task comes from the computation of the inner products $\gamma \sim \Braket{Q,Q}_e$ and $\alpha \sim \Braket{b,Q}_e$. This computation requires ${\cal O}(\ell_{\rm max}^3)$ operations in a brute-force approach and it is inevitable when obtaining $\alpha_n^{X_1 X_2 X_3}$, unless the starting bispectrum is already written in separable form; note, however, that this calculation is required only once, for a given shape. The expansion coefficients $\alpha$ can be computed and stored away for each shape under study, without constituting a bottleneck for the data estimation pipeline. Since the $Q$-basis is separable, we can always speed up the calculation of $\gamma_{np}^{X_1 X_2 X_3}$, by using the separable formula \cite{Smith:2006ud}:
\begin{eqnarray}
  \begin{split}
\gamma_{np}^{X_1 X_2 X_3} &= 8\pi^2
\int_{-1}^1 d\mu \, 
\zeta_{[i [i'}^{X_1} (\mu) 
\zeta_{j j'}^{X_2} (\mu) 
\zeta_{k] k']}^{X_3} (\mu) , \\
\zeta_{i i'}^X (\mu) 
&\equiv  \sum_{\ell} \frac{2\ell +1}{4\pi} \frac{q_i^X (\ell) q_{i'}^X (\ell)}{v_\ell^2} P_\ell(\mu) .
\end{split}
\label{eq:fastgamma}
\end{eqnarray}
The derivation of this formula is based on the following integral representation of the Wigner $3j$ symbols:
\begin{equation}
h_{\ell_1 \ell_2 \ell_3}^2
= \frac{(2\ell_1 +1)(2\ell_2 + 1)(2\ell_3 + 1)}{8 \pi }
\int_{-1}^1 d\mu \,
 P_{\ell_1}(\mu) P_{\ell_2}(\mu) P_{\ell_3}(\mu) , \label{eq:hh_open}
\end{equation} 
where $P_{\ell}(\mu)$ is the Legendre polynomial. Employing eq.~\eqref{eq:fastgamma} allows for a massive speed up of the algorithm, since it reduces the total number of operations by a factor $\ell_{\rm max}$.


\subsection{Separable estimator}\label{sec:est}

As firstly pointed out in ref.~\cite{Komatsu:2003iq}, if we can factorize the bispectrum, it is always possible to write the estimator \eqref{eq:estimator_2D} in separable form, by resorting on 
the identity (Gaunt integral):
\begin{equation}
  h_{\ell_1 \ell_2 \ell_3}
  \left(
  \begin{array}{ccc}
  \ell_1 & \ell_2 & \ell_3 \\
  m_1 & m_2 & m_3 
  \end{array}
  \right)
  = \int d^2 \hat{\bf n} \, Y_{\ell_1 m_1}(\hat{\bf n}) Y_{\ell_2 m_2}(\hat{\bf n}) Y_{\ell_3 m_3}(\hat{\bf n}) . \label{eq:Gaunt}
\end{equation}
If we consider our factorized decomposition for auto- and cross-bispectra, we can generalize the modal estimator derivation of ref.~\cite{Fergusson:2009nv}, obtaining that the joint auto- and cross-bispectrum estimator can be rewritten into an integral of the products of the following filtered maps [note how separability is made possible also by the choice of weights in eq.~\eqref{eq:expansion}]:
\begin{eqnarray}
  M_{i}^{X}(\hat{\bf n}) &=& \sum_{\ell m} 
\frac{q_i^X(\ell)}{v_\ell} \, 
\overline{a}_{\ell m}^X  Y_{\ell m}(\hat{\bf n}) , \label{eq:M_even} \\
 \overline{a}_{\ell m}^X &\equiv& \sqrt{C_{\ell}^{XX}} \sum_{X'} (C_\ell^{-1})^{XX'} a_{\ell m}^{X'} \label{eq:almbar} .
\end{eqnarray}
Using this expression, eq.~\eqref{eq:estimator_2D} reduces to 
\begin{eqnarray}
  {\cal E}
  &=& \frac{1}{6F} \sum_{X_i} \sum_{n}
  \alpha_n^{X_1 X_2 X_3} \beta_{n}^{X_1 X_2 X_3} \nonumber \\
  &=& \frac{1}{6F} \sum_{n} 
\left( \alpha_n^{YYY} \beta_{n}^{YYY} + 
 \alpha_n^{ZZZ} \beta_{n}^{ZZZ} + 
 3 \alpha_n^{YYZ} \beta_{n}^{YYZ} + 
 3 \alpha_n^{ZZY} \beta_{n}^{ZZY} 
 \right) , \label{eq:estimator_2D_mod}
\end{eqnarray}
where the new coefficients $\beta_n^{X_1 X_2 X_3} \equiv \beta_{{\rm cub} \, n}^{X_1 X_2 X_3} - \beta_{{\rm lin} \, n}^{X_1 X_2 X_3}$ contain all the information about the observed bispectra ${\cal B}_{\ell_1 \ell_2 \ell_3}^{X_1 X_2 X_3}$. Their cubic and linear parts are represented, respectively, as
\begin{eqnarray}
  \begin{split}
  \beta_{{\rm cub} \, ijk}^{X_1 X_2 X_3} &= 
\int d^2 \hat{\bf n} \, 
M_{[i}^{ X_1}(\hat{\bf n}) 
M_j^{X_2}(\hat{\bf n}) 
M_{k]}^{X_3}(\hat{\bf n}) , \\ 
\beta_{{\rm lin} \, ijk}^{X_1 X_2 X_3} &=  \int d^2 \hat{\bf n} \,  
\left[ 
\Braket{ M_{[i}^{X_1}(\hat{\bf n}) 
M_j^{X_2}(\hat{\bf n}) } 
  M_{k]}^{X_3}(\hat{\bf n})
+ \Braket{ M_{[j}^{X_2}(\hat{\bf n}) 
M_k^{X_3}(\hat{\bf n})  }
M_{i]}^{X_1}(\hat{\bf n}) \right. \\
& \left. \qquad\qquad 
+ \Braket{ M_{[k}^{X_3}(\hat{\bf n}) 
M_i^{X_1}(\hat{\bf n}) }
M_{j]}^{X_2}(\hat{\bf n}) 
\right] .
\end{split}
 \label{eq:betadef}
\end{eqnarray}
To summarize, the simplicity of the estimator in eq.~\eqref{eq:estimator_2D_mod} has been created by the linear transformation of the observed $a^X_{\ell m}$ described by eq.~\eqref{eq:almbar}.  It is this transformation which removed all the cross terms from the estimator.  This is what will allow us to simply estimate bispectra both jointly and individually using this method. This is in contrast to other methods like that in ref.~\cite{Fergusson:2014gea} that instead removes the cross terms in the estimator by orthonormalizing the $a^X_{\ell m}$ which entangles the individual bispectra so they can only be jointly constrained.  The two approaches can be simply related by:
\begin{equation}
  \left(
  \begin{array}{c}
    \overline{a}_{\ell m}^Y \\
    \overline{a}_{\ell m}^Z
  \end{array}
  \right)
  \equiv
  \left(
  \begin{array}{cc}
    1 \quad & \frac{-C_\ell^{YZ}}{\sqrt{C_\ell^{YY} C_\ell^{ZZ} - (C_\ell^{YZ})^2}}  \\
    0 \quad & \frac{\sqrt{C_\ell^{YY} C_\ell^{ZZ}}}{\sqrt{C_\ell^{YY} C_\ell^{ZZ} - (C_\ell^{YZ})^2}} 
  \end{array}
  \right)
  \left(
  \begin{array}{c}
    \hat{a}_{\ell m}^Y \\
    \hat{a}_{\ell m}^Z
  \end{array}
  \right)
\end{equation}
where $\hat{a}_{\ell m}^X \equiv \sum_{X'} L_\ell^{XX'} a_{\ell m}^{X'}$, with $L_\ell^{XX'}$ the rotational matrix \eqref{eq:L_def}, are the orthonormalized multipoles of ref.~\cite{Fergusson:2014gea}.

In order to get a complete expression for a fast estimator, we need to ensure that also the normalization can be computed fast. This requires fast numerical evaluation of the Fisher matrix for our expanded templates. If we deal only with auto-bispectra ($YYY$), then this is straightforward, since it is easy to verify, in such case, that:
\begin{equation}
 F_{YYY} = \frac{1}{6} \sum_{n = 0}^{n_{\rm max}} \left( \alpha_n^{R YYY} \right)^2 ,
\end{equation}
where $n_{\rm max}$ is the total number of basis templates used in the expansion. The situation becomes however more complex in the joint auto- and cross-bispectrum estimation case which we are considering here, since 
now the different bispectra are correlated, and we need to evaluate their full covariance. The straightforward, but slow approach would be to explicitly compute our approximating 
bispectrum template by summing over the modes ($b \sim \sum_n \alpha_n Q_n$). We can then insert the re-summed $b_{(e) \ell_1 \ell_2 \ell_3}^{X_1 X_2 X_3}$ into eq.~\eqref{eq:F_2D}  and numerically evaluate this expression. This would scale like ${\cal O}(\ell_{\rm max}^3)$, even when the input bispectrum shape is explicitly separable. Furthermore, the whole process needs to be repeated not just for any new shape, but also any time we change the noise properties in the input data, since the noise power spectrum appears in the weights. This is very inconvenient, especially when a large number of data validation and consistency checks are performed (as it was the case, for example, in the {\it Planck} primordial NG data validation campaign, for which the currently described pipeline was intensively used).  

Fortunately, the separability of the modal basis allows also in this case for a large speed up of the algorithm. We start by expanding the Fisher matrix~\eqref{eq:F_2D} in the $Q$-basis:
\begin{equation}
  F = \frac{1}{6} \sum_{X_i X_i'} \sum_{np} \alpha_{n}^{X_1 X_2 X_3}
  \tau_{np}^{X_1 X_2 X_3 X_1' X_2' X_3'}
  \alpha_{p}^{X_1' X_2' X_3'} ,
  \label{eq:F_2D_mod}
\end{equation}
where $\tau^{X_1 X_2 X_3 X_1' X_2' X_3'}_{n p}$ is a $8(n_{\rm max} + 1) \times 8(n_{\rm max} + 1)$ real symmetric matrix, reading 
\begin{equation}
\tau_{np}^{X_1 X_2 X_3 X_1' X_2' X_3'}  \equiv 
\Braket{Q_{n}^{X_1 X_2 X_3} Q_{p}^{X_1' X_2' X_3'}
\left[ \prod_{i = 1}^3 \sqrt{C_{\ell_i}^{X_i X_i}} (C_{\ell_i}^{-1})^{X_i X_i'} 
\sqrt{ C_{\ell_i}^{X_i' X_i'}  } \right] }_{e} . \label{eq:taudef}
\end{equation}
Exploiting eq.~\eqref{eq:hh_open}, we can then apply the same factorization trick as done in eq.~\eqref{eq:fastgamma}
\begin{eqnarray}
  \tau_{np}^{X_1 X_2 X_3 X_1' X_2' X_3'} 
&=& 8\pi^2 \int_{-1}^1 d\mu \, 
 {\cal Z}_{X_1' [i'}^{X_1 [i }(\mu) 
{\cal Z}_{X_2' j'}^{X_2 j}(\mu) 
{\cal Z}_{X_3' k']}^{X_3 k]}(\mu) \label{eq:tau_even_mod}  , \\
{\cal Z}_{X' i'}^{X i}(\mu) &\equiv& 
\sum_{\ell} \frac{2\ell +1}{4\pi} 
\frac{q_i^X(\ell) q_{i'}^{X'}(\ell)}{v_\ell^2} 
\sqrt{C_{\ell}^{X X}} (C_{\ell}^{-1})^{X X'} 
\sqrt{ C_{\ell}^{X' X'}}
 P_\ell(\mu). \label{eq:Z_even}  
\end{eqnarray}
The number of required numerical operations falls again by ${\cal O}(\ell_{\rm max})$ times, in comparison with the brute force computation. Note how we are computing 
here the covariance of the basis modes, which we indicated as $\tau_{np}$. This is completely model independent. The computation of $\tau_{np}$ can therefore always be carried efficiently, even for non-separable input bispectra. 
Finally, since $\tau_{np}$ represents essentially the mode covariance matrix between different bispectra, it also plays a key role when the estimator is used for full, model-independent bispectrum reconstruction, rather than for measuring $f_{\rm NL}$; see section~\ref{sec:reconstruction} for further explanations on this issue.

\subsection{Non-Gaussian simulations}

The procedure described in the previous section can be similarly applied to produce a factorized expression of the NG simulation formula, eq.~\eqref{eq:Alm_sim}, leading to:
\begin{equation}
A_{\ell m}^{X_1 X_2 X_3} 
= \frac{1}{6} 
 \sum_{X_i'} \sum_{ijk} \alpha_{ijk}^{X_1' X_2' X_3'} 
\int d^2 \hat{\bf n}  
\left[ 
  \Upsilon_{[i, \ell m}^{X_1 X_1'}(\hat{\bf n})
   G_{j}^{X_2 X_2'}(\hat{\bf n}) G_{ k]}^{X_3 X_3'}(\hat{\bf n}) 
  \right]^* , \label{eq:Alm_sim_even_mod}
\end{equation}
where
 \begin{eqnarray}
 \Upsilon_{i, \ell m}^{X X'}(\hat{\bf n})
&\equiv& Y_{\ell m} (\hat{\bf n}) L_\ell^{X X'} \frac{q_i^{X'}(\ell)}{v_\ell} \sqrt{ C_\ell^{X' X'}}  , \\
G_{i}^{X X'}(\hat{\bf n})
&\equiv& \sum_{\ell m} 
g_{\ell m}^{X} \, \Upsilon_{i, \ell m}^{X X'}(\hat{\bf n}) .
\end{eqnarray}
This allows creating a NG map through ${\cal O}(\ell_{\rm max}^4)$ operations.

\section{Spin-weighted modal estimator for parity-odd bispectra}\label{sec:odd}

In several important and interesting cases, the angular bispectrum takes values on a parity-odd domain:
\begin{equation}
  \ell_1 + \ell_2 + \ell_3 = {\rm odd} , \ \ \
  |\ell_1 - \ell_2| \leq \ell_3 \leq |\ell_1 + \ell_2| . \label{eq:domain_odd}
\end{equation}
This happens not only in presence of parity-breaking primordial models, but also e.g. when evaluating cross-bispectra between CMB temperature and polarization B-modes (TTB), in a parity-conserving Universe. It is therefore useful to extend the previously described algorithm for fast estimation, to include the parity-odd case.

The derivation procedure is essentially the same as in the parity-even case, aside from the use of a spin-weighted geometrical factor:
\begin{equation}
h_{\ell_1 \ell_2 \ell_3}^{x~ y~ z} 
\equiv \sqrt{\frac{(2\ell_1 + 1)(2\ell_2 + 1)(2\ell_3 + 1)}{4\pi}}
\left(
  \begin{array}{ccc}
  \ell_1 & \ell_2 & \ell_3 \\
  x & y & z 
  \end{array}
 \right).
\end{equation}
Here, the bispectrum templates obeying the parity-odd condition~\eqref{eq:domain_odd}, $B_{(o) \ell_1 \ell_2 \ell_3}^{X_1 X_2 X_3}$, are decomposed into  
\begin{equation}
B_{(o) \ell_1 \ell_2 \ell_3}^{X_1 X_2 X_3} 
= {}_{[x y z]}H_{\ell_1 \ell_2 \ell_3}^{X_1 X_2 X_3}
b_{(o) \ell_1 \ell_2 \ell_3}^{X_1 X_2 X_3},
\end{equation}
where $b_{(o) \ell_1 \ell_2 \ell_3}^{X_1 X_2 X_3}$ are the parity-odd reduced bispectra and  
\begin{eqnarray}
{}_{[x y z]} H_{\ell_1 \ell_2 \ell_3}^{YYY} &\equiv& 
\frac{1}{6} \left( h_{\ell_1 \ell_2 \ell_3}^{x ~y ~z} + 5~{\rm perms~in~}x,y,z \right) , \\ 
{}_{[x y z]} H_{\ell_1 \ell_2 \ell_3}^{YYZ} &\equiv& \frac{1}{2}\left(h_{\ell_1 \ell_2 \ell_3}^{x ~y ~z} + h_{\ell_1 \ell_2 \ell_3}^{y ~x ~z} \right) .
\end{eqnarray}
Hence $b_{(o) \ell_1 \ell_2 \ell_3}^{X_1 X_2 X_3} = b_{(o) \ell_2 \ell_1 \ell_3}^{X_2 X_1 X_3} = b_{(o) \ell_1 \ell_3 \ell_2}^{X_1 X_3 X_2} = b_{(o) \ell_3 \ell_1 \ell_2}^{X_3 X_1 X_2}= b_{(o) \ell_2 \ell_3 \ell_1}^{X_2 X_3 X_1} = b_{(o) \ell_3 \ell_2 \ell_1}^{X_3 X_2 X_1}$ holds. In the zero-spin limit, $x = y = z = 0$, ${}_{[x y z]} H_{\ell_1 \ell_2 \ell_3}^{X_1 X_2 X_3} = h_{\ell_1 \ell_2 \ell_3}^{0~0~0} = h_{\ell_1 \ell_2 \ell_3}$ just selects the even $\ell_1 + \ell_2 + \ell_3$ signal. In other words, in order to pick up all the nonvanishing signal satisfying both eq.~\eqref{eq:domain_odd} and $\ell_1, \ell_2, \ell_3 \geq 2$, we here have to choose nonzero $x$, $y$ and $z$. Then, the selection rules of $h_{\ell_1 \ell_2 \ell_3}^{x ~y ~z}$ force us to obey $x + y + z = 0$ and $-2 \leq x, y, z \leq  2$. There are some possible combinations, and one of them is $(x, y, z) = (1, 1, -2)$ \cite{Shiraishi:2014roa}.

In the next section, we will examine this case, and extend the parity-odd auto-bispectrum method developed in refs.~\cite{Shiraishi:2014roa,Shiraishi:2014ila}.

\subsection{Modal decomposition}

In analogy to parity-even bispectra, parity-odd reduced bispectra can be decomposed as
\begin{equation}
  \frac{v_{\ell_1} v_{\ell_2} v_{\ell_3} b_{(o) \ell_1 \ell_2 \ell_3}^{X_1 X_2 X_3}  }{i \sqrt{C_{\ell_1}^{X_1 X_1} C_{\ell_2}^{X_2 X_2} C_{\ell_3}^{X_3 X_3}} }  
= \sum_{n} \alpha_n^{X_1 X_2 X_3} Q_{n}^{X_1 X_2 X_3}(\ell_1, \ell_2, \ell_3) ,
\end{equation}
where we include the imaginary unit in the weighting, and $\alpha_n^{X_1 X_2 X_3} \in \mathbb{R}$, since $b_{(o) \ell_1 \ell_2 \ell_3}^{X_1 X_2 X_3}$ take pure imaginary values.

The inner product in odd $\ell_1 + \ell_2 + \ell_3$ space is defined with a spin-dependent weighting; namely,
\begin{equation}
  \Braket{f_{\ell_1 \ell_2 \ell_3}^{X_1 X_2 X_3 X_1' X_2' X_3'}}_{o}
  \equiv \sum_{\ell_1 + \ell_2 + \ell_3 = {\rm odd}} 
\frac{{}_{[x y z]}H_{\ell_1 \ell_2 \ell_3}^{X_1 X_2 X_3}}{v_{\ell_1} v_{\ell_2} v_{\ell_3}} 
\frac{{}_{[x y z]}H_{\ell_1 \ell_2 \ell_3}^{X_1' X_2' X_3'}}{v_{\ell_1} v_{\ell_2} v_{\ell_3}} 
f_{\ell_1 \ell_2 \ell_3}^{X_1 X_2 X_3 X_1' X_2' X_3'} . \label{eq:prod_odd_def}
\end{equation}
Unlike the parity-even case \eqref{eq:prod_even_def}, we here have to limit the summation range to $\ell_1 + \ell_2 + \ell_3 = {\rm odd}$ by hand. By employing it, the modal coefficients are expressed as
\begin{equation}
\alpha_n^{X_1 X_2 X_3} = \sum_p \left(\gamma^{X_1 X_2 X_2}\right)_{np}^{-1} 
\Braket{ \frac{v_{\ell_1} v_{\ell_2} v_{\ell_3} b_{(o) \ell_1 \ell_2 \ell_3}^{X_1 X_2 X_3} }{i \sqrt{C_{\ell_1}^{X_1 X_1} C_{\ell_2}^{X_2 X_2} C_{\ell_3}^{X_3 X_3}}} \, Q_p^{X_1 X_2 X_3} (\ell_1,\ell_2,\ell_3) }_{o} .  
\end{equation} 
The $\gamma_{np}^{X_1 X_2 X_3}$ matrix is defined and factorized as:
\begin{eqnarray} 
  \gamma_{np}^{X_1 X_2 X_3}
  &\equiv& \Braket{Q_{n}^{X_1 X_2 X_3}(\ell_1, \ell_2, \ell_3)  \, Q_{p}^{X_1 X_2 X_3}(\ell_1, \ell_2, \ell_3) }_{o} \nonumber \\
  &=& 8\pi^2 \int_{-1}^1 d\mu \sum_{a+b+c = o}
{}^{(a)}_{X_1}\zeta_{[x [i'}^{[-x [i }(\mu) \,
{}^{(b)}_{X_2}\zeta_{y j'}^{-y j} (\mu) \, 
{}^{(c)}_{X_3}\zeta_{z] k']}^{-z] k]} (\mu)
\label{eq:gamma_odd} ,
\end{eqnarray}
where the maps depending on spin and parity are given by
\begin{equation}
  {}^{(o/e)}_{\ \ \ X} \zeta_{x' i'}^{x i}(\mu)
  \equiv \sum_{\ell = {\rm odd/even}}\sqrt{\frac{2\ell +1}{4\pi}} 
\frac{q_i^X(\ell) q_{i'}^X(\ell)}{v_\ell^2} {}_{x}\lambda_{\ell x'}(\mu), 
\end{equation}
and $\sum_{a+b+c = o}$ denotes a summation over parity-odd pairs, i.e., $\sum_{a+b+c = o} {}^{(a)} \zeta {}^{(b)} \zeta {}^{(c)} \zeta = {}^{(o)} \zeta {}^{(e)} \zeta {}^{(e)}\zeta + {}^{(e)} \zeta {}^{(o)} \zeta {}^{(e)} \zeta + {}^{(e)} \zeta {}^{(e)} \zeta {}^{(o)} \zeta + {}^{(o)} \zeta {}^{(o)} \zeta {}^{(o)} \zeta$. 
To derive the above, we have used the following formula:  
\begin{equation}
h_{\ell_1 \ell_2 \ell_3}^{x~y~z} h_{\ell_1 \ell_2 \ell_3}^{x' y' z'}
= \sqrt{\pi(2\ell_1 +1)(2\ell_2 + 1)(2\ell_3 + 1)} \int_{-1}^1 d\mu \,
{}_{-x} \lambda_{\ell_1 x'}(\mu) {}_{-y} \lambda_{\ell_2 y'}(\mu) {}_{-z} \lambda_{\ell_3 z'}(\mu) , \label{eq:hh_spin_open}
\end{equation} 
where ${}_s \lambda_{\ell m}(\mu) \equiv {}_{s}Y_{\ell m}(\mu, \phi) e^{- i m \phi}$.

Similarly to the parity-even case, we can move to the orthonormal basis via the rotation~\eqref{eq:Q2R}. The decomposition into the orthonormal basis is written as
\begin{equation}
  \frac{v_{\ell_1} v_{\ell_2} v_{\ell_3}  b_{(o) \ell_1 \ell_2 \ell_3}^{X_1 X_2 X_3} }{i \sqrt{C_{\ell_1}^{X_1 X_1} C_{\ell_2}^{X_2 X_2} C_{\ell_3}^{X_3 X_3}} } = 
\sum_n \alpha_n^{R X_1 X_2 X_3} R_n^{X_1 X_2 X_3} (\ell_1, \ell_2, \ell_3) , 
  \end{equation}
where the coefficients, given by
\begin{equation}
\alpha_n^{R X_1 X_2 X_3} =
\Braket{ \frac{v_{\ell_1} v_{\ell_2} v_{\ell_3} b_{(o) \ell_1 \ell_2 \ell_3}^{X_1 X_2 X_3} }{i \sqrt{C_{\ell_1}^{X_1 X_1} C_{\ell_2}^{X_2 X_2} C_{\ell_3}^{X_3 X_3}}} \, R_n^{X_1 X_2 X_3} (\ell_1,\ell_2,\ell_3) }_{o} ,
\end{equation}
are related to $\alpha_n^{X_1 X_2 X_3}$ through eq.~\eqref{eq:aR2aQ}.

\subsection{Separable estimator}

The parity-odd estimator can be factorized in the same way as the parity-even case, only by replacing the Gaunt integral~\eqref{eq:Gaunt} with its spin-weighted version, reading
\begin{equation}
  h_{\ell_1 \ell_2 \ell_3}^{x~y~z}   \left(
  \begin{array}{ccc}
  \ell_1 & \ell_2 & \ell_3 \\
  m_1 & m_2 & m_3 
  \end{array}
  \right)
  = \int d^2 \hat{\bf n} \, {}_{-x}Y_{\ell_1 m_1}(\hat{\bf n}) {}_{-y} Y_{\ell_2 m_2}(\hat{\bf n}) {}_{-z} Y_{\ell_3 m_3}(\hat{\bf n}).  \label{eq:gaunt_spin}
\end{equation}
We can now generalize eqs.~\eqref{eq:M_even}~and~\eqref{eq:Z_even} to include 
information on spin and parity, obtaining new filtered maps:
\begin{eqnarray}
  {}^{(o/e)}M_{xi}^{X}(\hat{\bf n})
  &\equiv& \sum_{\ell = {\rm odd / even}} \sum_m 
  \frac{q_i^X(\ell)}{v_\ell} \,
  \overline{a}_{\ell m}^{X} \, {}_{x}Y_{\ell m}(\hat{\bf n}), \\
{}^{(o/e)} {\cal Z}_{X' x' i'}^{X x i}(\mu)
     &\equiv& \sum_{\ell = {\rm odd/even}}\sqrt{\frac{2\ell +1}{4\pi}} 
\frac{q_i^X(\ell) q_{i'}^{X'}(\ell)}{v_\ell^2} 
\sqrt{C_{\ell}^{X X}} (C_{\ell}^{-1})^{X X'} 
\sqrt{ C_{\ell}^{X' X'}}
{}_{x}\lambda_{\ell x'}(\mu) , 
\end{eqnarray}
where $\overline{a}_{\ell m}^X$ is given by eq.~\eqref{eq:almbar}. The parity-odd estimator can now be written exactly in the same form as eq.~\eqref{eq:estimator_2D_mod}; namely 
\begin{equation} 
  {\cal E}
  = \frac{\sum_{X_i} \sum_{n} \alpha_n^{X_1 X_2 X_3} \beta_{n}^{X_1 X_2 X_3}}{\sum_{X_i X_i'} \sum_{np} \alpha_{n}^{X_1 X_2 X_3} \tau_{np}^{X_1 X_2 X_3 X_1' X_2' X_3'}  \alpha_{p}^{X_1' X_2' X_3'} } , 
\end{equation}
however, the coefficients $\beta_{ijk}^{X_1 X_2 X_3} =
  \beta_{{\rm cub} \, ijk}^{X_1 X_2 X_3} - \beta_{{\rm lin} \, ijk}^{X_1 X_2 X_3}$ and the matrix $\tau_{np}^{X_1 X_2 X_3 X_1' X_2' X_3'} $ are now composed of the new filtered maps:
  \begin{eqnarray}
    \begin{split}
  \beta_{{\rm cub} \, ijk}^{X_1 X_2 X_3} &= \frac{1}{i} 
\int d^2 \hat{\bf n} \sum_{a+b+c = o} 
{}^{(a)} M_{[-x [i}^{X_1}(\hat{\bf n}) 
{}^{(b)} M_{-y j}^{X_2}(\hat{\bf n}) 
{}^{(c)} M_{ -z] k]}^{X_3}(\hat{\bf n}) , \\ 
\beta_{{\rm lin} \, ijk}^{X_1 X_2 X_3} &= \frac{1}{i}  \int d^2 \hat{\bf n} \sum_{a+b+c = o} 
\left[ 
\Braket{ {}^{(a)} M_{[-x [i}^{X_1}(\hat{\bf n}) 
{}^{(b)} M_{-y j}^{X_2}(\hat{\bf n}) } 
{}^{(c)} M_{-z] k]}^{X_3}(\hat{\bf n}) \right. \\
&\left.\qquad\qquad\qquad\qquad 
+ \Braket{ {}^{(a)} M_{[-y [j}^{X_2}(\hat{\bf n}) 
{}^{(b)} M_{-z k}^{X_3}(\hat{\bf n}) } 
{}^{(c)} M_{-x] i]}^{X_1}(\hat{\bf n})  \right. \\
&\left.\qquad\qquad\qquad\qquad 
+ \Braket{ {}^{(a)} M_{[-z [k}^{X_3}(\hat{\bf n}) 
{}^{(b)} M_{-x i}^{X_1}(\hat{\bf n}) } 
{}^{(c)} M_{-y] j]}^{X_2}(\hat{\bf n}) 
\right] ,
\end{split}
\label{eq:betadefodd}
\end{eqnarray}
and the mode expansion of the bispectrum covariance now reads:
\begin{eqnarray}
  \tau_{np}^{X_1 X_2 X_3 X_1' X_2' X_3'} &\equiv&
\Braket{Q_{n}^{X_1 X_2 X_3} Q_{p}^{X_1' X_2' X_3'}
\left[ \prod_{i = 1}^3 \sqrt{C_{\ell_i}^{X_i X_i}} (C^{-1}_{\ell_i})^{X_i X_i'} 
\sqrt{ C_{\ell_i}^{X_i' X_i'}  } \right] }_{o} \nonumber \\
  &=& 8\pi^2 \int_{-1}^1 d\mu \sum_{a+b+c = o} 
{}^{(a)}{\cal Z}_{X_1' [x [i' }^{X_1 [-x [i} (\mu) \,
{}^{(b)}{\cal Z}_{X_2' y j'}^{X_2 -y j} (\mu) \,
{}^{(c)}{\cal Z}_{X_3' z] k']  }^{X_3 -z] k]} (\mu) . \label{eq:tau_odd_mod}
\end{eqnarray}
This concludes the construction of our parity-odd fast estimator.

\subsection{Non-Gaussian simulations}

The algorithm to produce NG simulations can also be generalized to the parity-odd case. The NG part of the simulated $a_{\ell m}^{X}$, given by eq.~\eqref{eq:Alm_sim}, is now rewritten into the separable form:
\begin{eqnarray}
A_{\ell m}^{X_1 X_2 X_3} 
&=& \frac{i}{6} 
 \sum_{X_i'} \sum_{ijk} \alpha_{ijk}^{X_1' X_2' X_3'} 
 \int d^2 \hat{\bf n}
\Upsilon_{[-x[i, \ell m}^{X_1 X_1' *}(\hat{\bf n})   
\nonumber \\ 
&& 
\begin{cases}
  \left[{}^{(o)}G_{-y j}^{X_2 X_2'}(\hat{\bf n}) {}^{(o)}G_{-z] k]}^{X_3 X_3'}(\hat{\bf n})
  + {}^{(e)}G_{-y j}^{X_2 X_2'}(\hat{\bf n}) {}^{(e)}G_{-z] k]}^{X_3 X_3'}(\hat{\bf n}) 
\right]^*
 & : \ell = {\rm odd} 
  \\ 
  \left[ {}^{(o)}G_{-y j}^{X_2 X_2'}(\hat{\bf n}) {}^{(e)}G_{-z] k]}^{X_3 X_3'}(\hat{\bf n})
 + {}^{(e)}G_{-y j}^{X_2 X_2'}(\hat{\bf n}) {}^{(o)}G_{-z] k]}^{X_3 X_3'}(\hat{\bf n})
   \right]^* 
& : \ell = {\rm even} 
\end{cases}
, \label{eq:Alm_sim_odd_mod}
\end{eqnarray}
where 
\begin{eqnarray}
 \Upsilon_{xi, \ell m}^{X X'}(\hat{\bf n})
  &\equiv& {}_x Y_{\ell m} (\hat{\bf n}) 
 L_\ell^{X X'} \frac{q_i^{X'}(\ell) }{v_\ell} \sqrt{ C_\ell^{X' X'}} ,\\
{}^{ (o/e)}G_{x i}^{X X'}(\hat{\bf n})
&\equiv& \sum_{\ell = {\rm odd / even}} \sum_{m}  g_{\ell m}^{X} \,
\Upsilon_{xi, \ell m}^{X X'}(\hat{\bf n}) .
\end{eqnarray}
This enables fast numerical computation, as in the parity even case.

\section{General bispectrum estimation and bispectrum reconstruction}\label{sec:mixed}

So far we have described the implementation of a general modal pipeline for joint estimation of the $f_{\rm NL}$ bispectrum amplitude parameter, using all possible bispectra built from the combination of 
two weakly NG field, $Y$ and $Z$. The starting fast estimator expression, eq.~\eqref{eq:estimator_2D}, was firstly obtained by the authors of ref.~\cite{Yadav:2007rk}, who started from a $\chi^2$ statistic for the bispectrum, and managed to express the full covariance between different auto- and mixed-bispectra in factorizable form. This approach yields the cubic part of the estimator. The linear (``mean field'') term in $a_{\ell m}$ can then be computed by minimizing the variance of the cubic part, in presence of realistic isotropy-breaking effects, such as sky masking or an inhomogenous spatial noise distribution.

Since the derivation of the separable estimator formula, eq.~\eqref{eq:estimator_2D}, was only briefly sketched in ref.~\cite{Yadav:2007rk}, we provide here a detailed derivation, starting from a different, yet customary, approach, based on an Edgeworth expansion of the $a_{\ell m}$ likelihood in presence of mild NG. We show in particular that starting from the likelihood allows us to obtain directly the factorizable estimator expression, without further manipulations of the bispectrum covariance. We can then straightforwardly derive expression for the fast, factorized estimator of $f_{\rm NL}$ from a given cross-bispectrum -- say $B_{\ell_1 \ell_2 \ell_3}^{YYZ}$ -- alone (not jointly with other cross-bispectra and auto-bispectra). Finally, and importantly, this approach also allows deriving a formula for full reconstruction of auto- and cross-bispectra, in a model-independent way, and at no computational extra-cost; we will discuss this further in sections~\ref{sec:cross} and \ref{sec:reconstruction}.

The Edgeworth expansion of the $f_{\rm NL}$ likelihood, generalized from the auto-bispectrum case \cite{Babich:2005en}, is:
\begin{eqnarray}
P({\bf a}) &=& \left( 1 - \frac{1}{6} \sum_{X_i} \sum_{\ell_i m_i} \Braket{a_{\ell_1 m_1}^{X_1} a_{\ell_2 m_2}^{X_2} a_{\ell_3 m_3}^{X_3}} 
\frac{\partial}{\partial a_{\ell_1 m_1}^{X_1}} \frac{\partial}{\partial a_{\ell_2 m_2}^{X_2}} \frac{\partial}{\partial a_{\ell_3 m_3}^{X_3}} + \cdots \right) \nonumber \\ 
&& \frac{\exp\left(-\frac{1}{2} \sum_{X X'} \sum_{\ell m \ell' m'} a_{\ell m}^{X *} 
({\cal C}^{-1})_{\ell m \ell' m'}^{X X'} 
a_{\ell' m'}^{X'} \right)}{(2\pi)^{N_p/2} (det {\cal C})^{1/2}} \nonumber \\ 
&\equiv& P_0({\bf a}) + P_1({\bf a}) + \cdots , \label{eq:Edgew}
\end{eqnarray}
where ${\cal C}_{\ell m \ell' m'}^{X X'} \equiv \Braket{a_{\ell m}^{X *} a_{\ell' m'}^{X'}}$. Then, because of the weakness of NG,
\begin{equation}
\ln P({\bf a}) = \ln P_0({\bf a}) + \frac{P_1 ({\bf a})}{P_0 ({\bf a})} + \cdots
\label{eq:lnP}
\end{equation}
holds, where, one can find, by explicitly taking the derivatives:
\begin{eqnarray}
 \frac{P_1({\bf a})}{P_0({\bf a})} &=& \frac{1}{6} \sum_{X_i} \sum_{\ell_i m_i} 
 \Braket{a_{\ell_1 m_1}^{X_1} a_{\ell_2 m_2}^{X_2} a_{\ell_3 m_3}^{X_3}}
\sum_{X_i'} \sum_{\ell_i' m_i'} 
(C^{-1})_{\ell_1 m_1 \ell_1' m_1'}^{X_1 X_1'} 
(C^{-1})_{\ell_2 m_2 \ell_2' m_2'}^{X_2 X_2'} (C^{-1})_{\ell_3 m_3 \ell_3' m_3'}^{X_3 X_3'} 
\nonumber \\ 
&&
\left[ 
  a_{\ell_1' m_1'}^{X_1'} a_{\ell_2' m_2'}^{X_2'} a_{\ell_3' m_3'}^{X_3'}
- \left( a_{\ell_1' m_1'}^{X_1'} C_{\ell_2' m_2' \ell_3' m_3'}^{X_2' X_3'} + 2 \ {\rm perms} \right)
\right],  \label{eq:P1byP0}
\end{eqnarray}
with $C_{\ell m \ell' m'}^{X X'} \equiv \Braket{a_{\ell m}^{X} a_{\ell' m'}^{X'}} = (-1)^m {\cal C}_{\ell -m \ell' m'}^{X X'}$.

This expression is the starting point for the derivation of the estimators, described in the following. 

\subsection{Amplitude estimator}

We start by considering a general case in which each of the possible bispectra out of the combination of $Y$ and $Z$ has an independent amplitude, i.e., we now deal with the set of parameters $\{f_{\rm NL}^{YYY}, f_{\rm NL}^{ZZZ}, f_{\rm NL}^{YYZ}, f_{\rm NL}^{ZZY} \}$. The usual expression~\eqref{eq:estimator_2D} for the $f_{\rm NL}$ estimator is of course immediately derived from the results in this section, imposing that the four amplitudes are the same. The slightly more general case discussed here can be directly useful, for example, to test the presence of foreground residuals in primordial NG analysis of CMB temperature and polarization data. In principle, these residuals can in fact generate spurious bispectra with different amplitudes \cite{Coulton:2019bnz}. We note that, due to correlations between temperature and polarization data, a test of this kind would probably be robust only after preliminary projecting temperature out of the polarization component, in order to make the measured amplitudes independent. Moreover, the algebraic steps below, using independent amplitudes, can be straightforwardly generalized to obtain a (model-independent) mode- and bispectrum-reconstruction estimator, see section~\ref{sec:reconstruction}. 

An optimal, unbiased estimator ${\cal E}_I$ must satisfy the following conditions (the former defining unbiasedness, and the latter the saturation of the Cram\'er-Rao bound):
\begin{equation}
  \Braket{{\cal E}_I} = f_{\rm NL}^I \ \ \ \cap \ \ \
  \Braket{{\cal E}_I {\cal E}_J}
  = (F^{-1})_{IJ} , \label{eq:optimal_2D_YYY_YYZ} 
\end{equation}
where $I,J \in \{YYY, ZZZ, YYZ, ZZY \}$ and the $4 \times 4$ Fisher matrix is defined as 
\begin{equation}
  F_{IJ} \equiv \Braket{\frac{\partial \ln P({\bf a})}{\partial f_{\rm NL}^{I}}
  \frac{\partial \ln P({\bf a})}{\partial f_{\rm NL}^{J}} } .
\end{equation}
With eqs.~\eqref{eq:lnP}~and~\eqref{eq:P1byP0}, the score function is computed as
\begin{equation}
 \frac{\partial \ln P({\bf a})}{\partial f_{\rm NL}^{ I }} 
= \frac{1}{6} \sum_{X_1 X_2 X_3}^{\{ I \}} \sum_{X_i'} \sum_{\ell_i } 
(-1)^{\ell_1 + \ell_2 + \ell_3}
B_{\ell_1 \ell_2 \ell_3}^{X_1 X_2 X_3} 
(C_{\ell_1}^{-1})^{X_1 X_1'} 
(C_{\ell_2}^{-1})^{X_2 X_2'} (C_{\ell_3}^{-1})^{X_3 X_3'}
{\cal B}_{\ell_1 \ell_2 \ell_3}^{X_1' X_2' X_3'} , \label{eq:score}
\end{equation}
where $\displaystyle \sum_{X_1 X_2 X_3}^{\{ YYY \}} \equiv \sum_{X_1 X_2 X_3}^{YYY}$ and $\displaystyle \sum_{X_1 X_2 X_3}^{\{ YYZ \}} \equiv \sum_{X_1 X_2 X_3}^{\substack{ YYZ \\ YZY \\ ZYY}} $. For derivation, we have used the diagonal covariance matrix approximation:
\begin{equation}
  (C^{-1})_{\ell m \ell' m'}^{X X'} \simeq (C_\ell^{-1})^{X X'} (-1)^{m} \delta_{\ell, \ell'} \delta_{m, -m'}.
\end{equation}
Each element of the Fisher matrix is thus given by
\begin{equation}
  F_{I J} = \frac{1}{6}
\sum_{X_1 X_2 X_3}^{\{ I \}}
\sum_{X_1' X_2' X_3'}^{\{ J \}}
\sum_{\ell_i} (-1)^{\ell_1 + \ell_2 + \ell_3}  
B_{\ell_1 \ell_2 \ell_3}^{X_1 X_2 X_3}
(C_{\ell_1}^{-1})^{X_1 X_1'} (C_{\ell_2}^{-1})^{X_2 X_2'} (C_{\ell_3}^{-1})^{X_3 X_3'} 
B_{\ell_1 \ell_2 \ell_3}^{X_1' X_2' X_3'}. 
\end{equation}
Imposing the conditions \eqref{eq:optimal_2D_YYY_YYZ} then leads to the following expression for the general estimator:
\begin{eqnarray}
{\cal E}_I
&=& \sum_J^{\substack{YYY \\ ZZZ \\ YYZ \\ ZZY}} (F^{-1})_{IJ} \frac{\partial \ln P({\bf a})}{\partial f_{\rm NL}^J} \nonumber \\
&=& 
\frac{1}{6} \sum_J^{\substack{ YYY \\ ZZZ \\ YYZ \\ ZZY}} (F^{-1})_{IJ}
 \sum_{X_1 X_2 X_3}^{\{J\}} \sum_{X_i'} \sum_{\ell_i } 
(-1)^{\ell_1 + \ell_2 + \ell_3}
B_{\ell_1 \ell_2 \ell_3}^{X_1 X_2 X_3} \nonumber \\ 
&& (C_{\ell_1}^{-1})^{X_1 X_1'} 
(C_{\ell_2}^{-1})^{X_2 X_2'} (C_{\ell_3}^{-1})^{X_3 X_3'}
{\cal B}_{\ell_1 \ell_2 \ell_3}^{X_1' X_2' X_3'}. \label{eq:estimator_2D_YYY_YYZ}
\end{eqnarray}

We can now apply the modal decomposition to the expression above, leading to the final form for the modal estimator 
\begin{eqnarray}
{\cal E}_I  &=& \frac{1}{6} \sum_J^{\substack{ YYY \\ ZZZ \\ YYZ \\ ZZY}} (F^{-1})_{IJ}
\sum_{X_1 X_2 X_3}^{\{J\}}
\sum_n \alpha_n^{X_1 X_2 X_3} \beta_n^{X_1 X_2 X_3} \nonumber \\
&=& \frac{1}{6} \sum_n
\left[ (F^{-1})_{I, YYY} \, \alpha_n^{YYY} \beta_n^{YYY}
  + (F^{-1})_{I, ZZZ} \, \alpha_n^{ZZZ} \beta_n^{ZZZ} \right. \nonumber \\ 
&& \left. \qquad + 3 (F^{-1})_{I, YYZ} \, \alpha_n^{YYZ} \beta_n^{YYZ}
  + 3 (F^{-1})_{I, ZZY} \, \alpha_n^{ZZY} \beta_n^{ZZY}
  \right] , \label{eq:estimator_2D_YYY_YYZ_mod} \\
F_{IJ} &=& \frac{1}{6} \sum_{X_1 X_2 X_3}^{\{ I \}}
\sum_{X_1' X_2' X_3'}^{\{ J \}}
\sum_{np} \alpha_n^{X_1 X_2 X_3} \tau_{np}^{X_1 X_2 X_3 X_1' X_2' X_3'} \alpha_p^{X_1' X_2' X_3'} , 
\end{eqnarray}
where $\alpha_n^{X_1 X_2 X_3}$, $\beta_n^{X_1 X_2 X_3}$ and $\tau_{np}^{X_1 X_2 X_3 X_1' X_2' X_3'}$ for even and odd $\ell_1 + \ell_2 + \ell_3$ modes are computed as described in sections~\ref{sec:even} and \ref{sec:odd}, respectively.
Notice that this expression is not the same as eq.~\eqref{eq:estimator_2D_mod}, due to different normalization factors. Of course it reduces to eq.~\eqref{eq:estimator_2D_mod} when $f_{\rm NL}^{YYY} = f_{\rm NL}^{ZZZ} = f_{\rm NL}^{YYZ} = f_{\rm NL}^{ZZY}$.

\subsection{Separable estimator for $f_{\rm NL}^{YYZ}$}\label{sec:cross}

Here we consider the estimation of $f_{\rm NL}^{YYZ}$ alone. Obviously, eqs.~\eqref{eq:estimator_2D_YYY_YYZ}~and~\eqref{eq:estimator_2D_YYY_YYZ_mod} directly yield the $f_{\rm NL}^{YYZ}$ estimator for models in which the other three bispectra vanish, i.e. $f_{\rm NL}^{YYY} = f_{\rm NL}^{ZZZ} = f_{\rm NL}^{ZZY} = 0$ (for such situations see e.g., refs.~\cite{Bartolo:2015fqz,Shiraishi:2016hjd,Meerburg:2016ecv,Domenech:2017kno,Bartolo:2018elp}). The optimal estimator, obeying
\begin{equation}
  \Braket{{\cal E}_{YYZ}} = f_{\rm NL}^{YYZ} \ \ \ \cap \ \ \
  \Braket{{\cal E}_{YYZ}^2 } = \frac{1}{F_{YYZ,YYZ}} , \label{eq:optimal_YYZ_1D}
\end{equation}
takes then the expected form:
\begin{eqnarray}
  {\cal E}_{YYZ}
  &=& \frac{1}{F_{YYZ, YYZ}} \frac{\partial \ln P({\bf a})}{\partial f_{\rm NL}^{YYZ}}  \nonumber \\
&=&  \frac{1}{6 F_{YYZ,YYZ} }
 \sum_{X_1 X_2 X_3}^{\{YYZ\}} \sum_{X_i'} \sum_{\ell_i } 
(-1)^{\ell_1 + \ell_2 + \ell_3}
B_{\ell_1 \ell_2 \ell_3}^{X_1 X_2 X_3} \nonumber \\ 
&&
(C_{\ell_1}^{-1})^{X_1 X_1'} 
(C_{\ell_2}^{-1})^{X_2 X_2'} (C_{\ell_3}^{-1})^{X_3 X_3'}
{\cal B}_{\ell_1 \ell_2 \ell_3}^{X_1' X_2' X_3'}. \label{eq:estimator_1D_YYZ}
\end{eqnarray}
Via the modal decomposition, the estimator reduces to:
\begin{equation}
{\cal E}_{YYZ} = \frac{1}{2 F_{YYZ, YYZ}} \sum_n \alpha_n^{YYZ} \beta_n^{YYZ}. \label{eq:estimator_1D_YYZ_mod} 
\end{equation}

\subsection{Bispectrum reconstruction}\label{sec:reconstruction}

In a previous work~\cite{Fergusson:2009nv} it was shown that the modal estimator does not just directly allows us to estimate the amplitude $f_{\rm NL}$, but also to reconstruct the full auto-bispectrum of the map. To this purpose, it suffices to consider e.g. the estimated $\beta^{YYY}$ coefficients from eqs.~\eqref{eq:betadef}~and~\eqref{eq:betadefodd}, and build the linear combination: 
\begin{eqnarray}
  {\hat{b}}^{YYY}_{\ell_1 \ell_2 \ell_3}
  &=&  \frac{\sqrt{C_{\ell_1}^{YY}C_{\ell_2}^{YY}C_{\ell_3}^{YY}}}{v_{\ell_1} v_{\ell_2} v_{\ell_3}} \sum_n \left(\sum_p \left( \gamma^{YYY} \right)^{-1}_{np} \beta_p^{YYY} \right) Q_{n}^{YYY}(\ell_1, \ell_2, \ell_3) \nonumber \\ 
&& \times 
  \begin{cases}
1 & :\ell_1 + \ell_2 + \ell_3 = {\rm even} \\
i & :\ell_1 + \ell_2 + \ell_3 = {\rm odd}
  \end{cases}
  . 
\end{eqnarray}
This can be generalized to the mixed bispectrum case. Therefore, as it is always the case with the modal approach, even if we started with an amplitude $f_{\rm NL}$ estimator, what we obtain is actually also a general method for full bispectrum reconstruction (with a finite resolution, imposed by the truncation of the modal expansion series, as it would be for any binning scheme).
The bispectrum reconstruction formula for the full auto- and cross-bispectrum case can be obtained by simply taking the expected value of the $\beta$ coefficients, defined in eqs.~\eqref{eq:betadef}~and~\eqref{eq:betadefodd}. This yields:
\begin{eqnarray}
\hat{b}_{\ell_1 \ell_2 \ell_3}^{X_1 X_2 X_3}   
&=& 
\frac{\sqrt{C_{\ell_1}^{X_1 X_1} C_{\ell_2}^{X_2 X_2} C_{\ell_3}^{X_3 X_3}} } {v_{\ell_1} v_{\ell_2} v_{\ell_3}  }
\sum_{n}
\left( \sum_p 
\sum_{X_i'} (\tau^{-1})^{X_1 X_2 X_3 X_1' X_2' X_3'}_{n p} 
\beta_p^{X_1' X_2' X_3'} \right) \nonumber \\ 
&& Q_{n}^{X_1 X_2 X_3}(\ell_1, \ell_2, \ell_3)
\times
  \begin{cases}
1 & :\ell_1 + \ell_2 + \ell_3 = {\rm even} \\
i & :\ell_1 + \ell_2 + \ell_3 = {\rm odd}
  \end{cases}
, \label{eq:reconstruction}
\end{eqnarray}
where the $\tau $ matrix was given by eqs.~\eqref{eq:tau_even_mod}~and~\eqref{eq:tau_odd_mod}. Note that our previous derivation of the $f_{\rm NL}$ estimator, in the case with four independent amplitudes, can also be used to obtain this reconstruction expression, via estimation of the expansion coefficients $\alpha$. To get the optimal $\hat{\alpha}$ estimator, simply consider eq.~\eqref{eq:P1byP0} and replace $\braket{ a_{\ell_1 m_1}^{X_1} a_{\ell_2 m_2}^{X_2} a_{\ell_3 m_3}^{X_3} }$ with the modal expansion:
\begin{eqnarray}
  \Braket{a^{X_1}_{\ell_1 m_1} a^{X_2}_{\ell_2 m_2} a^{X_3}_{\ell_3 m_3}}
  &=&  
  \frac{\sqrt{C_{\ell_1}^{X_1 X_1} C_{\ell_2}^{X_2 X_2} C_{\ell_3}^{X_3 X_3}}}{v_{\ell_1} v_{\ell_2} v_{\ell_3}}   \sum_n \alpha^{X_1 X_2 X_3}_n Q^{X_1 X_2 X_3}_n(\ell_1,\ell_2,\ell_3) \nonumber \\
  && 
  \left(
  \begin{array}{ccc}
  \ell_1 & \ell_2 & \ell_3 \\
  m_1 & m_2 & m_3 
  \end{array}
  \right)
  \times
  \begin{cases}
    h_{\ell_1 \ell_2 \ell_3} & : \ell_1 + \ell_2 + \ell_3 = {\rm even} \\
    i \, {}_{[x y z]}H_{\ell_1 \ell_2 \ell_3}^{X_1 X_2 X_3} & : \ell_1 + \ell_2 + \ell_3 = {\rm odd}
  \end{cases}
  . \label{eq:exprec}
\end{eqnarray}
Then, maximize with respect to the mode amplitudes $\alpha^{X_1 X_2 X_3}_n$, following completely analogous algebraic steps to what done for $f_{\rm NL}$ in the previous sections. The minimum variance estimate $\hat{\alpha}_n^ {X_1 X_2 X_3}$ can now be re-inserted in eq.~\eqref{eq:exprec} to obtain eq.~\eqref{eq:reconstruction}, which is then the optimal modal bispectrum reconstruction formula. The approach is of course valid both in the parity-even and parity-odd case.

\section{Pipeline implementation and validation}\label{sec:tests}

In this section, we test the modal estimators developed above via the measurements of CMB temperature and E-mode polarization bispectra, i.e., $B_{\ell_1 \ell_2 \ell_3}^{TTT}$, $B_{\ell_1 \ell_2 \ell_3}^{EEE}$, $B_{\ell_1 \ell_2 \ell_3}^{TTE}$ and $B_{\ell_1 \ell_2 \ell_3}^{EET}$, for both parity-even and parity-odd cases.

For theoretical templates of the bispectra, we adopt the parity-even one from the usual scalar-mode equilateral NG \cite{Creminelli:2005hu} and the parity-odd one from the tensor-mode equilateral NG \cite{Shiraishi:2013kxa,Shiraishi:2014ila,Ade:2015ava}. For these two cases, we produce NG simulations, using eqs.~\eqref{eq:Alm_sim_even_mod}~and~\eqref{eq:Alm_sim_odd_mod} respectively. 

We assume the amplitude of the bispectra to be $f_{\rm NL}^{\rm equil} = 300$ and $f_{\rm NL}^{\rm tens} = 2000$. We start by computing the averages of our estimators from $50$ NG simulations, considering both joint and independent estimation. These are described respectively by eq.~\eqref{eq:estimator_2D} -- with $\beta$ adapted to either the parity-even or parity-odd case -- and by eqs.~\eqref{eq:estimator_2D_YYY_YYZ_mod}~and~\eqref{eq:estimator_1D_YYZ_mod}.

We then extract the standard deviations, $\sqrt{\Braket{{\cal E}^2}}$ and $\sqrt{\Braket{{\cal E}_I^2}}$, from a separate set of 500 Gaussian realizations. For simplicity, we assume an ideal full-sky noiseless survey and hence all types of secondary uncertainties due to, e.g., finite beam widths, sky-cuts by masking, instrumental noises and foreground contaminations need not be taken into account (we note however again that this pipeline was thoroughly tested and applied to the primordial NG analysis of {\em Planck} data, where all these complications are obviously present).

In both cases, the modal decomposition is truncated when we reach a $95\%$ level of correlation between the original and the expanded shape. To speed up the computation, the multipole ranges are restricted to $2 \leq \ell_T \leq 500$ and $35 \leq \ell_E \leq 500$ in the parity-even analysis, and $2 \leq \ell_T, \ell_E \leq 100$ in the parity-odd one.

The averages and the standard deviations of each estimator are summarized in tables~\ref{tab:fNL} and \ref{tab:err}, respectively. We can see how the input $f_{\rm NL}$ values and the Cram\'er-Rao bounds are correctly recovered, both in the joint and independent estimation case; thus, we conclude that our estimators work properly.%
\footnote{One may feel at a glance that, in the parity-odd tensor NG case, $\braket{{\cal E}_{TTT}}$ and $\braket{{\cal E}_{EEE}}$ are a bit far from the input $f_{\rm NL}$ values, however, note that the error bars are also relatively large as shown in table~\ref{tab:err}, therefore the results are completely consistent.}

\begin{table}[t]
\begin{center}
\begin{tabular}{|c||c|c||c|c|} \hline
  & \multicolumn{2}{c||}{equilateral NG} & \multicolumn{2}{c|}{tensor NG} \\
  & \multicolumn{2}{c||}{($\ell_1 + \ell_2 + \ell_3 = \rm even$)} & \multicolumn{2}{c|}{($\ell_1 + \ell_2 + \ell_3 = \rm odd$)} \\ \cline{2-5}
  & $f_{\rm NL} = 300$ & $f_{\rm NL}^{TTE} = 300$ & $f_{\rm NL} = 2000$ & $f_{\rm NL}^{TTE} = 2000$ \\ \hline\hline
  $\Braket{{\cal E}}$ \eqref{eq:estimator_2D_mod} & {\bf 292} & 123 & {\bf 2030} & 1450 \\ \hline
  $\Braket{{\cal E}_{TTT}}$ \eqref{eq:estimator_2D_YYY_YYZ_mod} & {\bf 301} & 1.66 & {\bf 3430} & 1510 \\
  $\Braket{{\cal E}_{EEE}}$ \eqref{eq:estimator_2D_YYY_YYZ_mod} & {\bf 290} & $-1.67$ & {\bf 1670} & $-234$  \\   
  $\Braket{{\cal E}_{TTE}}$ \eqref{eq:estimator_2D_YYY_YYZ_mod} & {\bf 290} & {\bf 298} & {\bf 2030} & {\bf 2020} \\ 
  $\Braket{{\cal E}_{EET}}$ \eqref{eq:estimator_2D_YYY_YYZ_mod} & {\bf 297} & 3.75 & {\bf 2000} & $-21.7$ \\ \hline
  $\Braket{{\cal E}_{TTE}}$ \eqref{eq:estimator_1D_YYZ_mod} & 64.3 & {\bf 295} & 2090 & {\bf 2030} \\ \hline
  \end{tabular}
\end{center}
\caption{Averages of three types of estimators \eqref{eq:estimator_2D_mod}, \eqref{eq:estimator_2D_YYY_YYZ_mod} and \eqref{eq:estimator_1D_YYZ_mod} computed from 50 CMB maps sourced by the parity-even scalar equilateral NG (left two columns) and the parity-odd tensor equilateral one (right two columns). The maps used in the columns ``$f_{\rm NL} = A$'' and ``$f_{\rm NL}^{TTE} = A$'' are created assuming $(f_{\rm NL}^{TTT}, f_{\rm NL}^{EEE}, f_{\rm NL}^{TTE}, f_{\rm NL}^{EET}) = (A, A, A, A)$ and $(0, 0, A, 0)$, respectively. If the estimators are optimal, the bold numbers are expected to reach $A$ 
  .} 
\label{tab:fNL}
\begin{center}
  \begin{tabular}{|c||c||c|} \hline
    & equilateral NG & tensor NG  \\ 
    & ($\ell_1 + \ell_2 + \ell_3 = \rm even$) & ($\ell_1 + \ell_2 + \ell_3 = \rm odd$) \\ \hline\hline
  $\sqrt{\Braket{{\cal E}^2}}$ \eqref{eq:estimator_2D_mod} & 53.3 (50.5) & 351 (356) \\ \hline
  $\sqrt{\Braket{{\cal E}_{TTT}^2}}$ \eqref{eq:estimator_2D_YYY_YYZ_mod} & 96.5 (89.8) & 6520 (6300)  \\
  $\sqrt{\Braket{{\cal E}_{EEE}^2}}$ \eqref{eq:estimator_2D_YYY_YYZ_mod} & 78.2 (77.3) & 2220 (2250)  \\  
  $\sqrt{\Braket{{\cal E}_{TTE}^2}}$ \eqref{eq:estimator_2D_YYY_YYZ_mod} & 65.8 (62.2) & 426 (431)  \\ 
  $\sqrt{\Braket{{\cal E}_{EET}^2}}$ \eqref{eq:estimator_2D_YYY_YYZ_mod} & 67.1 (65.7) & 715 (706) \\ \hline
  $\sqrt{\Braket{{\cal E}_{TTE}^2}}$ \eqref{eq:estimator_1D_YYZ_mod} & 39.3 (39.3) & 421 (427) \\ \hline
  \end{tabular}
\end{center}
\caption{Standard deviations of three types of estimators \eqref{eq:estimator_2D_mod}, \eqref{eq:estimator_2D_YYY_YYZ_mod} and \eqref{eq:estimator_1D_YYZ_mod} computed from 500 Gaussian CMB maps. If the estimators are optimal, the results should converge to the Cram\'er-Rao bounds computed from the Fisher matrices shown in the brackets.} 
\label{tab:err}
\end{table}

\section{Conclusions}\label{sec:conclusions}

In this paper, we have proposed and described in detail a general modal methodology for reconstructing and estimating amplitudes of auto- and cross-angular bispectra, considering two correlated weakly-NG fields. As usual in the modal bispectrum approach, any theoretically-predicted shape is decomposed using suitable, separable modal basis templates, which are then fit to the data. The freedom in the choice of the 
modal basis, and the possibility to use it to get factorizable expansions of any input shape, make this approach very versatile.

In a previous study \cite{Fergusson:2014gea}, the modal decomposition was performed after the four input bispectrum templates ($YYY$, $ZZZ$, $YYZ$ and $ZZY$) were orthogonalized via a proper rotational transformation. In contrast, in the present approach, original bispectrum templates are decomposed as they are without rotation, and their covariance is dealt with later, again via a separable modal expansion. Both methodologies, of course, yield the same results when jointly constraining the amplitude of the four bispectra, $f_{\rm NL}$, given in eq.~\eqref{eq:fNL_def}. The orthogonalized method is slightly faster for joint estimation, whereas the method developed here directly enables us to 
independently reconstruct and estimate amplitudes of single cross-bispectra, $f_{\rm NL}^{YYZ}$ or $f_{\rm NL}^{ZZY}$ \eqref{eq:estimator_1D_YYZ_mod}, since those are decomposed from the start without rotation. The methodology presented here generalizes our previous work also by including the odd $\ell_1 + \ell_2 + \ell_3$ domain. This significantly extends the range of shapes which can be investigated. The pipeline presented here is an ideal starting point to include B-mode information in future modal CMB bispectrum analysis, since some bispectra have in this case odd-parity and we are explicitly interested in independent estimation of the TTB cross-bispectrum. \cite{Kamionkowski:2010rb,Shiraishi:2011st,Shiraishi:2013vha,Shiraishi:2013kxa,Meerburg:2016ecv,Domenech:2017kno,Bartolo:2018elp}. 
However, while $TTB$ estimation will be one of the main future applications for our pipeline, it is by no means the only one. Other interesting cross-bispectra arise for example in NG studies of spectral distortion anisotropies \cite{Bartolo:2015fqz,Shiraishi:2016hjd}. More in general, this methodology can be a flexible starting point for estimation of angular auto- and cross-bispectra for general random fields, including e.g. galaxy clustering and gravitational lensing studies.
We showed in section~\ref{sec:tests} some applications and tests of our pipeline, including NG maps generation and bispectrum estimation, with both parity-even and parity-odd examples. We also extensively validated and applied this same pipeline, as one of the main methods for primordial NG analysis of {\em Planck} temperature and polarization data \cite{Ade:2015ava}.
Further applications will be discussed in future studies.

\acknowledgments

M.\,S. was supported by JSPS Grant-in-Aid for Research Activity Start-up Grant Number 17H07319 and JSPS Grant-in-Aid for Early-Career Scientists Grant Number 19K14718. Numerical computations by M.\,S. were in part carried out on Cray XC50 at Center for Computational Astrophysics, National Astronomical Observatory of Japan. Part of this work was undertaken on the STFC COSMOS@DiRAC HPC Facility at the University of Cambridge, funded by UK BIS NEI grants





\bibliography{paper}
\end{document}